\documentclass[11pt]{article}
\usepackage[dvips]{graphicx}

\date{\today}

\begin{document}
\begin{titlepage}
\begin{center}
{\large\bf {Soft annealing: A new approach to difficult computational
  problems }\\[.3in]
 Nicolas Sourlas} \\
   Laboratoire de Physique Th\'eorique de l'Ecole Normale Sup\'erieure
         \footnote{Unit{\'e} Mixte de Recherche du CNRS et de
         l'Ecole Normale Sup\'erieure, associ\'ee
         \`a l'Universit\'e  Pierre et Marie Curie, PARIS VI.},\\
  24 rue Lhomond, 75231 Paris CEDEX 05, France
\end{center}
\vskip .15in
\centerline{\bf ABSTRACT}

\begin{quotation}

I propose a new method to study computationally difficult problems. 
I consider a new system, larger than the one I want to simulate. 
The original 
system is recovered by imposing constrains on the large system.
I simulate the large system with the hard constrains replaced 
by soft constrains. I illustrate the method in the case of the 
ferromagnetic Ising model and in the case the three dimensional 
spin-glass  model. 
I show that in both models 
the phases of the soft problem have  the same  properties 
as the phases of the original model  and  that the softened model 
belongs to the same 
universality class as the original one.

I show that correlation times are much shorter in the larger  
soft constrained system and that it is computationally advantageous to 
study it instead of the original system. 

This method is quite general and can be applied to many other systems. 

\vskip 0.5cm
\noindent


\end{quotation}
\end{titlepage}

It is well known that there exists a large class of systems which are
particularly hard to study both analytically and numerically. Such
systems are found in condensed matter physics (spin glasses, random 
field models and more generally  disordered systems) but also in 
combinatorial optimization problems or in communication theory (error 
correcting codes). The common future of those systems is the existence
of a large number of local minima, separated by large barriers. 
The simulation algorithm gets trapped in those minima.

\begin{figure}[b]
\includegraphics[width=12cm]{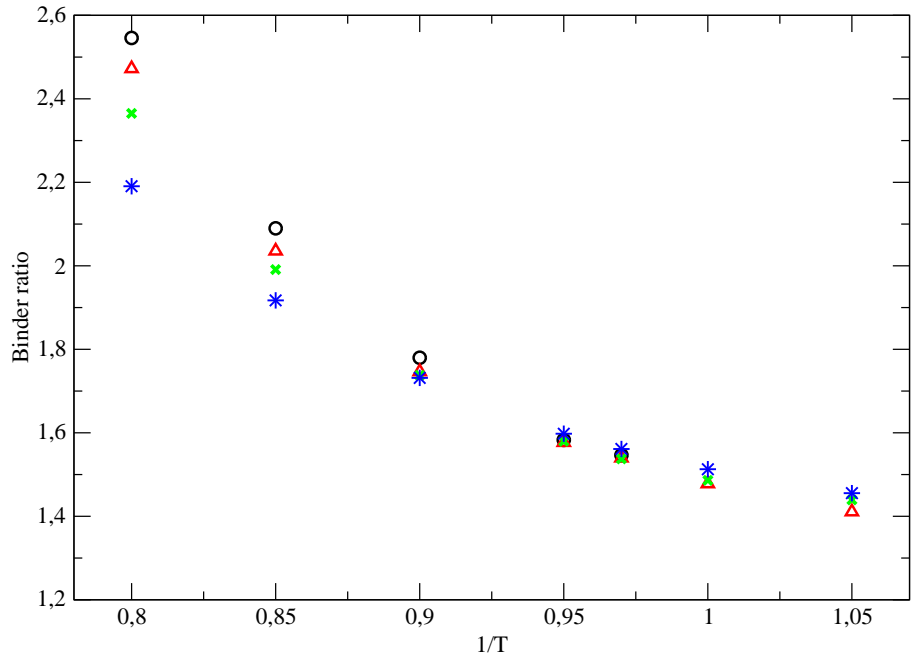}
\smallskip
\caption{Binder ratio $r$ plotted as a function of $\beta = 1/T $.
Black circles represent data for linear size $L=16$, red triangles for $L=14$, 
green $x$'s for $L=12$ and blue stars for $L=8$.
\label{fig1}}
\end{figure}

In this paper I propose a new method, I call soft annealing, in which 
I enlarge the space on which the original system is defined. The original 
system is recovered by imposing constrains on the large system.
I propose to simulate the large system with the hard constrains
replaced by soft constrains.  By enlarging the space on which the
problem is defined, it is hoped that one can get around the barriers 
and speed up the  dynamics of the algorithm, while retaining all the 
essential properties of the system. This is known to be 
the case for error 
correcting codes\cite{S}.  I will show that this is also the case for 
spin glasses and that it is  
advantageous to study the large system. 

I will illustrate the method 
with the example of spin models in three dimensional cubic lattices, 
but the method is more general. 

Consider a spin model described by the Hamiltonian 

$$  -H^0  =  \sum_{x,y,z = 1}^{L} J^x (x,y,z)  \sigma(x,y,z) \sigma(x+1,y,z) + 
J^y (x,y,z)  \sigma(x,y,z) \sigma(x,y+1,z) $$
$$ +  J^z (x,y,z)  \sigma(x,y,z) \sigma(x,y,z+1) $$

 where the $ \sigma $'s are Ising spins on a cubic lattice 
of linear size $L$. 

Now consider the following new  Hamiltonian: 
$$  -H^{new}  =  \sum_{x,y,z=1}^{L} J^x (x,y,z)  \sigma^{x}(x,y,z)
 \sigma^{x}(x+1,y,z) +  
J^y (x,y,z)  \sigma^{y}(x,y,z) \sigma^{y}(x,y+1,z) $$ 
$$ + J^z (x,y,z)  \sigma^{z}(x,y,z) \sigma^{z}(x,y,z+1) + $$
$$ u ( \sigma^{x}(x,y,z) 
 \sigma^{y}(x,y,z) + 
 \sigma^{x}(x,y,z)  \sigma^{z}(x,y,z) + 
 \sigma^{y}(x,y,z)  \sigma^{z}(x,y,z) )  $$

\begin{figure}[t]
\includegraphics[width=12cm]{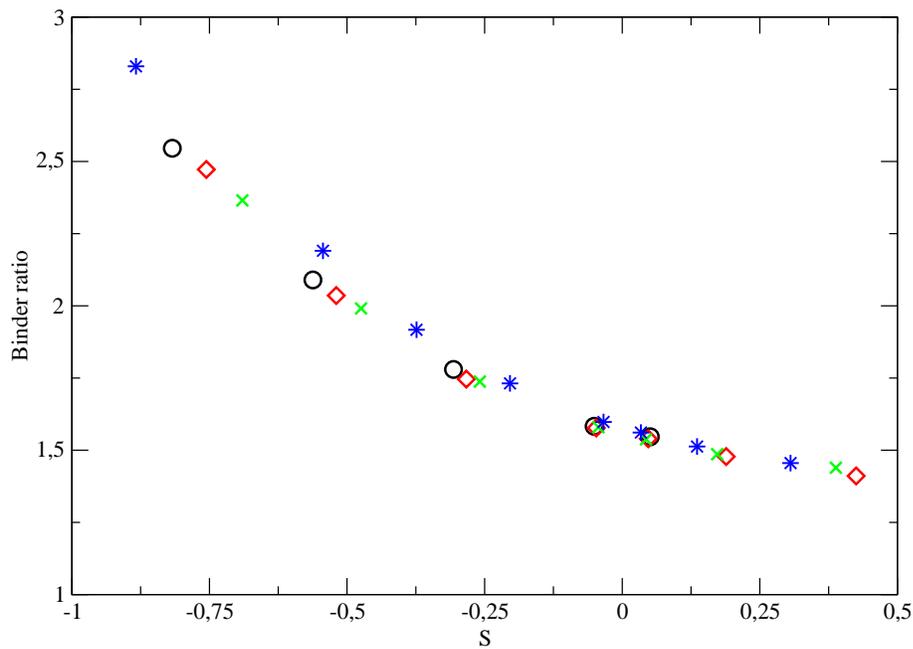}
\smallskip
\caption{Binder ratio $r$ plotted versus the scaling variable 
$s$ (see  text).  
Black circles represent data for linear size $L=16$, red squares for $L=14$, 
green $x$'s for $L=12$ and blue stars for $L=8$.
\label{fig2}}
\end{figure}

$ H^{new} $ contains three times more  spins than $ H^0 $. 
To every $ \sigma(x,y,z) $ of the original Hamiltonian $H^0$ correspond 
the three spins $  \sigma^{x}(x,y,z) $, $  \sigma^{y}(x,y,z) $ and 
$  \sigma^{z}(x,y,z) $. 
$  \sigma^{x}(x,y,z) $ is coupled to its near neighboring spins  
in the direction $x$, $  \sigma^{y}(x,y,z) $ in the direction $y$ 
and $  \sigma^{z}(x,y,z) $ in the direction $z$. The ferromagnetic
coupling $u$ couples together the three types of spin on 
every lattice site. For $ u = 0 $ $ H^{new} $ reduces to $3 L^2 $ 
decoupled one dimensional chains. For $u \to \infty $,  
 $  \sigma^{x}(x,y,z) =  \sigma^{y}(x,y,z) =
  \sigma^{z}(x,y,z) $ and $ H^{new} $ reduces to $ H^0 $.
I will call the model described by $ H^{new} $ the soft constrained model 
and the original  model described by $ H^0 $ the hard  constrained model. 

If  $ H^0 $ has a phase transition at $  \beta = 1/T = \beta_c $,
 we expect the phase diagramme of 
$ H^{new} $ to be two dimensional. For small $u$, $ H^{new} $  
is essentially one dimensional and there is no phase transition. 
For $  \beta  < \beta_c $ there is again no phase transition, 
irrespectively of the value of $ u $. 
In general, for $  \beta  < \beta_c $, there will be a line in the 
$ \beta = 1/T $, $ v= u \beta $  plane, separating the high temperature
from the low temperature phase. We expect the points along this critical 
line to be on the same universality class, i.e. the critical exponents 
 to remain the same, and equal to their value in the model described 
by $ H^0 $ i.e. the  $u \to \infty $ limit. 

First I verified that all this is true in the case of the
ferromagnetic Ising model in 3 dimensions. It is known that in 
this case  $\beta_c \sim  .22 $ . 
I simulated $ H^{new} $ for $\beta =.25 $ and different values of 
$ v $. I found a ferromagnetic to paramagnetic phase transition for 
$ v = v_c =.794 \pm .005 $. By finite size scaling I checked that the 
critical exponents $ \nu $ and the magnetic susceptibility exponent 
 are compatible with the best  
known values for the 3 dimensional Ising model.

 
Much more interesting is the case of the spin glass model 
because it is well known to be computationally hard. 
I simulated the Edwards Anderson model on a cubic lattice 
with periodic boundary conditions. The couplings $J $  are 
independent random variables taking the values $ \pm 1 $ with equal 
probability. This model has been studied a lot 
(for a review see reference\cite{MPR}). It undergoes a 
phase transition for $\beta =.90 $ and the value of the critical 
exponent $\nu $ is  $\nu = 1.7 \pm .3 $\cite{KY}. 

I studied the soft version of this model for different values of 
$ \beta $, keeping the ratio $ u =  v / \beta  $ fixed to 
$ u  =.75 $ and for the lattice sizes 
$ L= 8, \ 12, \ 14 $ and $ L= 16 $. For every size I simulated 
1280 realizations of the couplings. 
For every realization of the couplings I simulated two copies 
$ \sigma $ and $\tau $ and studied the probability distribution 
of the overlap $ q$\cite{P}  
\begin{equation} \label{q}
q = ( { 1 \over { 3 L^3 } } )  \sum_{x,y,z} ( \sigma^{x}(x,y,z)
\tau^{x}(x,y,z) + \sigma^{y}(x,y,z) \tau^{y}(x,y,z) +  
 \sigma^{z}(x,y,z) \tau^{z}(x,y,z) ) 
\end{equation}

I measured the Binder ratio $ r(\beta,L) =  {\overline {< q^4 >} }  
/ (  {\overline { < q^2 > } } )^2      $ 
 where, as usually, $ \overline{ a } $ 
 is the average of $a$ over the coupling samples, 
and the results are shown in figure 1.
The data for the different sizes cross at $\beta =.96 \pm .05 $. 
This shows that there is a phase transition for $ \beta = \beta_c = 
.96 \pm .05 $. 
Figure 2 shows the same data plotted versus the scaling 
variable $s = (\beta -\beta_c) L^{1/\nu} $ with $ \nu = 1.7 $, 
the exponent of the Edwards-Anderson model\cite{KY}. 
We see that the 
data for the different sizes collapse on the same curve except maybe for 
the $L=8$ data which seem to be slightly apart, which  
 indicates corrections to scaling for $L=8$. 
The value $r_c$ of $r$ at  $ \beta = \beta_c $ is 
an universal quantity, sometimes called the renormalized 
coupling constant. We find $r_c = 1.53 \pm .03 $. For the 
original model $ r_c \sim 1.5  $\cite{KY}. 
So our results are compatible with both the exponent $\nu $ and the
 renormalized coupling constant $r_c $ taking the same value in the 
soft model and in the original Edwards-Anderson model. We conclude 
that the hypothesis that the 
two models belong to the same universality class is supported by 
our data.

\begin{figure}[t]
\includegraphics[width=14cm]{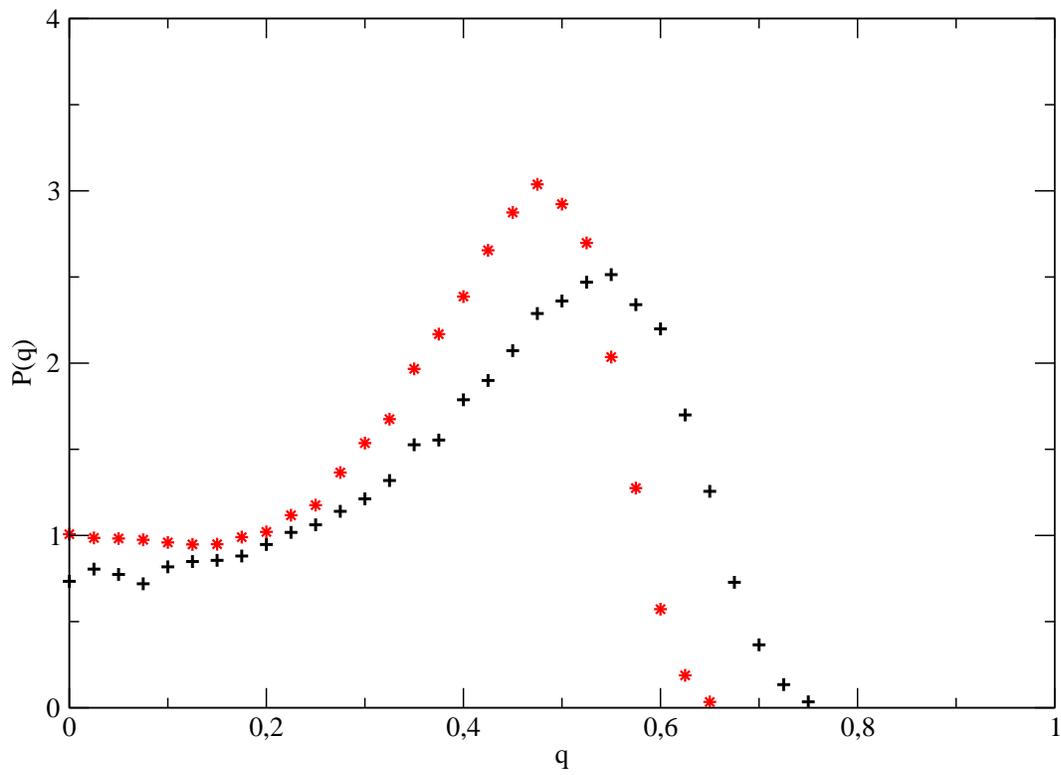}
\smallskip
\caption{Average probability distribution $P(q)$ of the overlaps $q$ 
for $\beta =1.05 $ and sizes $ L= 14 $ (stars in red) and $L = 8$ 
(crosses in black).
\label{fig3}}
\end{figure}

Figure 3 shows the probability distribution of the overlaps $P(q)$
averaged over the samples for $ \beta =1.05 $ and for the two sizes 
$L = 8 $ and $L=14$. The similarity with the  $P(q)$ of the hard
constrained model\cite{KY} is striking. $P(q)$ has a peack at a 
value of $q$ we call $q_{EA}$. As the volume increases, the peack 
is sharper and  $q_{EA}$ decreases, as in the original 
Edwards Anderson model. 
I measured also the link-link overlap $q_e$ 
of two real replicas $ \sigma $ and $\tau $,
 $ q_e = (1/3 L^3) \sum_{i,j} \sigma_i \sigma_j  \tau_i 
\tau_j $ (the sum runs over the nearest neighbor sites of the 
lattice) and the 
joined probability distribution, $P(q,q_e)$ 
averaged over the samples.
 I found that $P(q,q_e)$ is very peacked around a line 
$ q_e = e_0 + e_1 q^2 + e_2 q^4  $, with $ e_0 \sim .54 $, $ e_1  \sim
 .10 $ and 
$ e_2  \sim .30 $. The $ e_i $'s are to a  good approximation $L$ 
independent. I measured the variance 
$ w^2 = {\overline { < (q_{e} -e_{0} - e_{1 } q^{2} -
 e_{2 } q^{4})^2 > } } $.
I found for $\beta = 1.05 $  $ w^2 = .0004 $ 
for $ L= 8 $, $ w^2 = .00015 $ for  $L = 12 $,
  $ w^2 = .0001 $ for  $L = 14 $, and  $ w^2 = .00008 $ for  $L = 16 $,
i.e. when the volume 
increases $P(q,q_e)$ goes to a delta function 
 $P(q,q_e) = P(q) \delta ( q_e - e_0 + e_1 q^2 + e_2 q^4 ) $ 
 This is the case in the infinite range model and 
in the case of the hard constrained 
 model\cite{CPPS1,CPPS2,MAPA}. 
We conclude that the hard and soft constrained models belong to the 
same universality class and that their low temperature phases are
extremely similar. 

We expect the soft model to be easier to simulate numerically. 
To verify this one usually  measures the spin autocorrelation functions, 
averaged over the samples,  
$$ C^{h} (t)  = (1/L^3 ) \sum_{i} {\overline{ < \sigma_{i} (t_{0}) 
\sigma_{i} (t_{0}+t) > } } $$  and $$ C^{s} (t)  = (1 /3 L^3 ) 
{ \overline { < \sum_{i} \sigma_{i}^x  (t_{0}) 
\sigma_{i}^x  (t_{0}+t) +  \sigma_{i}^y  (t_{0}) \sigma_{i}^y (t_{0}+t)
 + \sigma_{i}^z  (t_{0}) \sigma_{i}^z (t_{0}+t)> } }  $$ 
 for the hard and 
soft constrained models respectively, where $ t $ is the number of 
Monte Carlo sweeps over the lattice. I compared $ C^{h} (t) $ and 
$ C^{s} (t) $ each one measured at the critical value of $\beta $,
$ \beta_{c}^{h} = .90 $ for $ C^{h} (t) $ and $ \beta_{c}^{s} = .97 $ 
for $ C^{s} (t) $.
They are plotted in figure 4. 

There are three curves in figure 4.
\begin{figure}[t]
\includegraphics[width=12cm, angle=270]{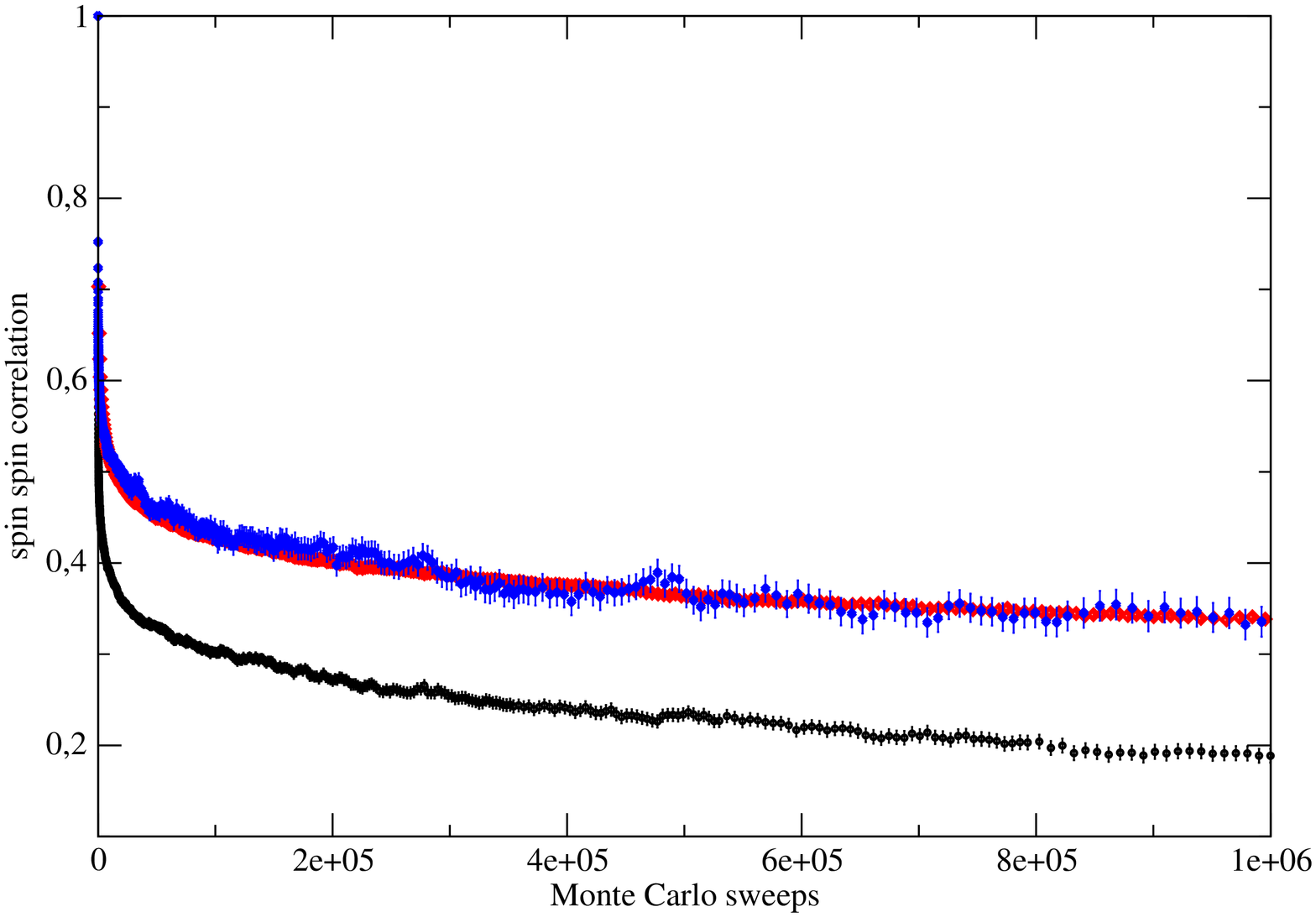}
\smallskip
\caption{Spin-spin autocorrelations $ C^{h} (t) $ and $ C^{s} (t) $ 
 at $ \beta^{h}_c $ and $ \beta^{s}_c $ (see text) 
 as a function of the number of the lattice Monte Carlo 
sweeps.
\label{fig4}}
\end{figure}

The lower curve (in black) corresponds to $ C^{s} (t) $. 
It is difficult to distinguish between  the two other curves 
because they fall on top of each other. 
One of the two, the curve in 
blue,  corresponds to $ C^{h} (t) $. We see that $ C^{s} (t) $ 
indeed decays much faster than $ C^{h} (t) $. In order to make 
the comparison more quantitative, I plotted also in figure 4 
 $ C^{s} (wt) $ (the curve in red), i.e. I rescaled the 
``time'' in $ C^{s} (t) $ by an appropriated scaling factor $w$. 
We see that, with this rescaling, 
 $ C^{s} (wt)$ and $ C^{h} (t)$ can be made to coincide. 

 The scaling factor $w$ was chosen as follows. 
$ C^{s} (t)$ and $ C^{h} (t)$ are decreasing functions of $t$ 
and $ C^{s} (0) =  C^{h} (0)= 1$. 
 I measure 
the number of lattice sweeps $ t^{h} ( c, \beta) $ and  
$ t^{s} ( c,\beta) $  needed for  $ C^{s} $ and  $ C^{h} $ 
fall below the value $ c $, i.e. 
 $ C^{h} ( t^{h} ( c, \beta)) =c $ 
and $ C^{s} ( t^{s} ( c, \beta)) =c $. For $ c = .40 $ and $ L = 12 $ 
I found that the time ratio 
$ r_t^{critical} (c = .40)  =  t^{h} ( .40 , \beta_{c}^{h}) /  
t^{s} ( .40 , \beta_{c}^{s}) 
\sim  28 $. In figure 4 I chose $ w =   r_t^{critical} (c = .40) $. 
 We see that 
with this rescaling of the Monte Carlo time, the 
spin autocorrelation functions coincide for all 
 $ 1 \le t \le10^6 $,   
 i.e. one has to perform  28 times more lattice 
sweeps for the hard constrained system in order for the 
spins to de-correlate the same amount as for the soft system. 
In order to study the size dependence of this difference  of the 
correlation times, I measured $ r_t^{critical} (c = .40) $ 
for $L=8$ and $L =16 $.
I found that  $ r_t^{critical} (c = .40) = 7.7 $ 
for $ L = 8 $ and $ r_t^{critical} (c = .40) = 39 $ for $ L= 16 $,   
 i.e. $ r_t^{critical} $ (and therefore the gain in simulation time) 
strongly increases with lattice size. 
It is not clear yet whether  $ r_t^{critical} $ obeys some kind 
of finite size scaling. 

I conclude that, by softening the 
constrains, the gain in computer time is large, despite the fact that 
the number of spins in $ H^{new}$ is larger by a factor of three. 
This gain icreases with the size of the system. 

The method of softening the constrains presented here is obviously 
not unique. For example one could consider 
$$ H' = \sum_{i,j} J_{i,j} X{i,j} + u \sum_{plaquettes}  X_{i,j} 
X_{j,k} X_{k,l} X_{l,i} $$
The term proportional to $u$ runs over all the plaquettes, i.e. 
the elementary squares, of the lattice. 
$  X_{i,j} $ corresponds to $ \sigma_i \sigma_j $ of the original 
Hamiltonian and the coupling $ u $ imposes the constrain 
$  X_{i,j} X_{j,k} X_{k,l} X_{l,i} =
\sigma_i \sigma_j \sigma_j \sigma_k \sigma_k \sigma_l \sigma_l \sigma_i = 1 $
 One of the advantages of the  method used in this paper is that 
all the techniques which speed up simulations like multi-spin 
coding and  parallel tempering can easily be implemented in the 
soft model. 

I have illustrated the method with the notoriously difficult 
example of the three dimensional spin glass model. It would 
be interesting to apply this method also to other difficult 
optimization problems. 

I would like to thank Andrea Montanari for several discussions.

 \end{document}